\documentclass[10pt,preprint]{sigplanconf}

\pdfoutput=1

\usepackage{amsmath}                        
\usepackage{amssymb}                        
\usepackage{booktabs}                       
\usepackage{caption}                        
\usepackage[usenames,dvipsnames]{color}     
\usepackage{datetime}                       
\usepackage[T1]{fontenc}                    
\usepackage[pdfborder={0 0 0}]{hyperref}    
\usepackage[utf8]{inputenc}                 
\usepackage{listings}                       
\usepackage{mathptmx}                       
\usepackage{tabu}                           
\usepackage{tikz}                           
\usepackage{url}                            
\usepackage{xspace}                         

\hypersetup{
  colorlinks,
  linkcolor={red!50!black},
  citecolor={blue!50!black},
  urlcolor={blue!80!black}
}

\definecolor{mygray}{rgb}{0.7,0.7,0.7}

\lstdefinelanguage{Go}{
  language=C,
  morekeywords={func,go,select,forever},
  literate={+}{{$+$}}1
           {-}{{$-$}}1
           {/}{{$/$}}1
           {*}{{$*$}}1
           {>}{{$>$}}1
           {<}{{$<$}}1
           {>=}{{$\geq$}}2
           {<=}{{$\leq$}}2
           {->}{{$\rightarrow$}}2
           {<-}{{$\leftarrow$}}2
           {//}{{//}}2
}

\newcommand{\blade}{\textsc{Blade}\xspace}

\copyrightyear{2015} 

\date{}

\authorinfo{David Terei\and Amit Levy}
           {Stanford University}
           {\{dterei,alevy\}@cs.stanford.edu}

\begin{document}

\title{Blade~: A Data Center Garbage Collector}
\maketitle

\begin{abstract}

An increasing number of high-performance distributed systems are written in
garbage collected languages. This removes a large class of harmful bugs from
these systems. However, it also introduces high tail-latency do to garbage
collection pause times.
We address this problem through a new technique of garbage collection avoidance
which we call \blade. \blade\ is an API between the collector and application
developer that allows developers to leverage existing failure recovery
mechanisms in distributed systems to coordinate collection and bound the
latency impact.
We describe \blade\ and implement it for the Go programming language. We also
investigate two different systems that utilize \blade, a HTTP load-balancer and
the Raft consensus algorithm. For the load-balancer, we eliminate any latency
introduced by the garbage collector, for Raft, we bound the latency impact to a
single network round-trip, (\(48\,\mu s\) in our setup). In both cases, latency
at the tail using \blade\ is up to three orders of magnitude better.

\end{abstract}

\section{Introduction}
\label{sec:intro}

Recently, there has been an increasing push for low-latency at the tail in
distributed systems~\cite{tail-at-scale, ramcloud2, lowlatency-now}.  This has
arisen from the needs of modern data center applications which consist of
hundreds of software services, deployed across thousands of machines. For
example, a single Facebook page load can involve fetching hundreds of results
from their distributed caching layer~\cite{fb-memcache-scaling}, while a Bing
search consists of 15 stages and involves thousands of servers in some of
them~\cite{speeding-req-resp}. These applications require latency in
microseconds with tight tail guarantees.

Recent work addressed this at the operating system and networking
layer~\cite{ix, arrakis, chronos, less-is-more}. However, this is only half the
picture. Increasingly, application developers choose to build distributed
systems in garbage collected lagnuages. For example, a large number of
distributed systems are written in Java~\cite{hdfs,spark,zookeeper}, and
Go~\cite{golang,gousers,go-dl.google.com,go-youtube}. Garbage collected
languages are attractive because manual memory management is extremely
bug-prone~\cite{buffer-overflow, undangle}.

\begin{figure}[t]
  \centering
  \hspace*{-5mm}
  \includegraphics[width=0.5\textwidth]{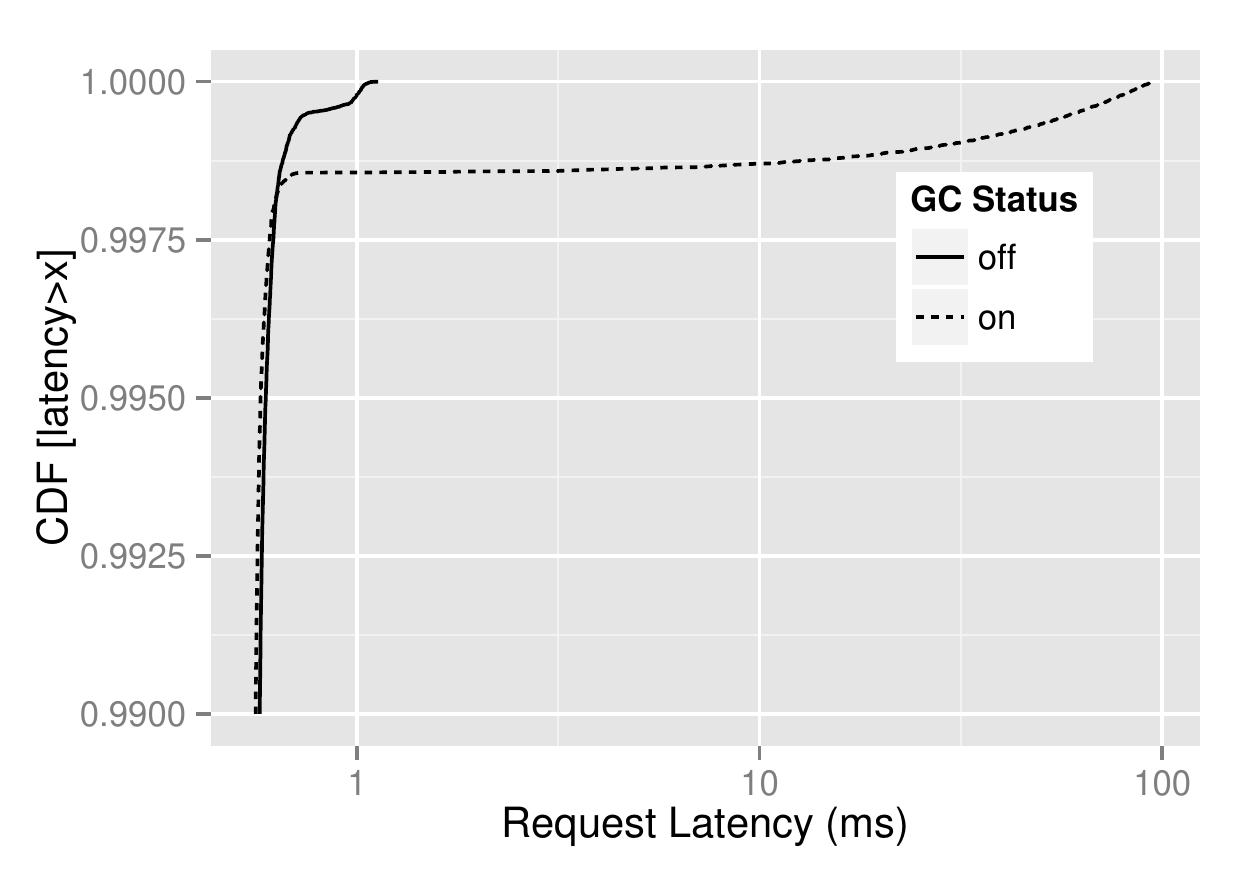}
  \caption{\label{fig:raft-gc}
    CDF of request latency for a ZooKeeper-like~\cite{zookeeper} replicated
    key-value store using the Raft~\cite{raft} consensus algorithm written in
    Go. System was configured with 3 nodes and 10 clients generating a total of
    250 requests-per-second (3:1 get/set ratio) over 10 minutes. A parallel,
    stop-the-world (STW) mark-sweep collector was used, with a heap size of
    \(500\,\text{MB}\) for a \(200\,\text{MB}\) working set.
  }
\end{figure}

Unfortunately, garbage collection introduces high tail-latencies due to long
pause times. For example, Figure~\ref{fig:raft-gc} shows the impact of garbage
collection on the tail-latency of one such distributed system. While many
distributed systems require average and tail-latencies in microseconds, garbage
collection pause times can range from milliseconds for small workloads to
seconds for large workloads.

Moreover, dealing with pause times in the application is hard. The impact of
garbage collection is often unpredictable during development and difficult to
debug once deployed. First, from the programmers perspective, garbage
collection can occur at any point during execution. Second, performance can
vary greatly from system to system, or even over the life-time of a single
system~\cite{fitzgerald-profile,soman-dynamic}. Finally, tuning the garbage
collector of a deployed system is hard because performance is workload
dependent. As a result, users must continually adjust run-time system
parameters (e.g., generation sizes) based on production workloads.

None of the current approaches to garbage collection are suitable for this new
set of requirements where the \(99.9\text{th}\) percentile matters. On one
side, language implementers attempt to build faster
collectors~\cite{compressor-gc, c4-gc, schism-gc, references-gc}. However they
are generally concerned with average case behaviour and optimising across a
large set of use-cases~\cite{dacapo}. As such, pause times at the tail are
still too long. Moreover, the effort to build better collectors must be
replicated for each language runtime. On the other side, developers deploying
such systems in production may turn off the garbage collector all
together\footnote{Generational garbage collectors often allow users only to
disable collection for the old generation, but this just serves to delay, not
eliminate, memory exhaustion}, or switch to manual memory management, giving up
the productivity gains of memory-safe languages~\cite{comparing-java}.

We propose a new approach to building distributed systems in garbage collected
languages, called \blade, that gives control over tail-latency back to the
programmer.
Instead of attempting to minimise pause times, distributed systems should treat
pause times as a frequent, but predictable \emph{failures}.
\blade\ is an interface to the run-time system that allows programmers to
participate in the decision to pause for collection, customising the collection
policy to their system.

\blade's simple API allows systems builders to:
\begin{enumerate}
  \item eliminate garbage collection related latency

  \item by leveraging system-specific failure recovery mechanisms to mask pause
    times,

  \item and model the performance impact of garbage collection without knowledge
    of the production workload.
\end{enumerate}

In this paper, we describe and evaluate the \blade\ API\@. We implemented
\blade\ for the Go programming language and used it to eliminate garbage
collection related tail-latency in two different distributed systems.
The first system is a cluster of web application servers behind a
load-balancer, and the second is the Raft~\cite{raft} consensus algorithm.

We compare \blade\ in both systems against the default Go garbage collector,
and the optimal solution for performance of no garbage collection at all. For
the HTTP cluster, \blade\ completely eliminates any latency impact on requests
caused by garbage collection, while for the Raft consensus algorithm, it bounds
the latency impact to a single extra network RTT (\(48\,\mu s\) in our
experimental setup). In end-to-end tests, this matches the performance of the
optimal system with no GC\@.

The rest of the paper is organized as follows. In Section~\ref{sec:background},
we motivate the problem and explain why existing solutions do not work. In
Section~\ref{sec:design} we outline \blade. In Section~\ref{sec:apps} we
explore two end-to-end distributed systems that use \blade\ and in
Section~\ref{sec:eval} we evaluate both systems. In
Section~\ref{sec:discussion} we discuss the results and limitations, while in
Section~\ref{sec:related} we describe related work. Finally, we conclude in
Section~\ref{sec:conclusion}.

\section{Background}
\label{sec:background}

\subsection{Data Center Performance Today}

The performance demands of applications running in data centers are changing
significantly. To enable rich interactions between services without impacting
the overall latency experienced by users, average latencies must be in the few
tens or low hundreds of microseconds~\cite{save-datacenter,lowlatency-now}.
Because a single user request may touch hundreds of servers, the long tail of
the latency distribution we must also consider~\cite{tail-at-scale,
speeding-req-resp}, with each service node ideally providing tight bounds on
even the \(99.9\text{th}\) percentile request latency.

Today, most commercial Memcached deployments provision each server so that the
99th percentile latency does not exceed \(500\,\mu s\)~\cite{leverich-qos}.
Recent academic results such as the \textsc{ix} operating system can run
Memcached with 99th percentile latencies of under \(100\,\mu s\) at
peak~\cite{ix}. The MICA key-value store can achieve 70 million
requests-per-second with tail-latencies of \(43\mu s\)~\cite{mica}.  Current
research projects such as RAMCloud~\cite{ramcloud1, ramcloud2} are targeting
\(10\,\mu s\) or lower RPC latencies.

\subsection{State of Garbage Collection}

We give a brief overview of garbage collection and the trade-offs for the main
approaches. Garbage collectors deal with two major concerns: finding and
recovering unused memory, and dealing with heap fragmentation, often by
relocating live objects. We will look at four collector designs: stop-the-world
(STW), concurrent, real-time and reference counting.

\paragraph{Stop-the-world}
Stop-the-world collectors are the oldest, simplest and highest throughput
collectors available~\cite{stw-gc-scale}. A STW GC works by first completely
stopping the application, then starting from a root set of pointers (registers,
stacks, global variables) traces out the applications live set. Objects are
either marked as live, or relocated to deal with fragmentation. Next, the
application can be resumed and any unmarked objects added to the free list.

Their simplicity and high-throughput make them common. For example, Go, Ruby,
and Oracle's JVM (by default) use STW collectors. The downside is that
pause times are proportional to the number of live pointers in the heap.  As a
result, state-of-the-art STW collectors can have pause times of
\(10\)--\(40\,\text{ms}\) per GB of heap~\cite{stw-gc-scale}.

\paragraph{Concurrent}
Concurrent collectors attempt to reduce the pause time caused by STW collectors
by enabling the GC to run concurrently with application threads. They achieve
this by using techniques such as read and write barriers to detect and fix
concurrent modifications to the heap while tracing live data and/or relocating
objects. For example, a common approach to concurrent tracing is to use write
barriers, either through inline code or virtual memory protection, whereby any
modification to the heap will enter a slow path handler that adds the pointer
to the list of pointers to trace~\cite{mapping-gc, compressor-gc, c4-gc}. For
handing concurrent relocation of objects to reduce fragmentation, either
Brook's style read barrier~\cite{brooks-gc} are used, where all pointer
dereferences check to see if the object has been replaced with a forwarding
pointer pointing to the objects new location, or, direct access
barriers~\cite{baker-realtime-gc, pauseless-gc, c4-gc}, where a read barrier is
used to fix up any pointer to point to the new location as the pointer is read
from the heap.

Because of these techniques, pause times for the best concurrent collectors are
measured in the few milliseconds~\cite{c4-gc, collie-gc, azul-white-paper1}.
However, concurrent collecters have lower-throughput, higher implementation
complexity and edge cases that still require GC pauses. First, concurrent
collectors reduce application throughput between \(10\)--\(40\%\) and increase
memory usage by \(20\%\) compared to STW
collectors~\cite{mapping-gc,g1-gc,stw-gc-scale}. This is due to the overhead of
handling barriers, forwarding pointers and synchronization between the GC
threads and application. Second, most concurrent collectors have corner cases
that trigger long pauses. For example, STW pauses are often used to start or
end collector phases~\cite{compressor-gc, mapping-gc}, the amount of work that
can occur in a slow-path for an allocation or barrier is variable and often
unbounded~\cite{compressor-gc, real-gc-survey}, and high-allocation rates can
cause the application to outpace the collector and pause~\cite{schism-gc}.
Finally, concurrent collectors are incredibly complex. Oracle's JVM, for
example, has two concurrent collectors, CMS and G1, but both have pause times
in the hundreds of milliseconds due to significant stages of their collection
cycle being STW\@.

\paragraph{Real-time}
Garbage collectors designed for real-time systems take the approaches of
concurrent collectors even further, many offering the ability to bound pause
times. The best collectors can achieve bounds in the tens of
microseconds~\cite{real-gc-survey, schism-gc}, however doing so comes at a high
throughput cost ranging from \(30\%\)--\(100\%\) overheads, and generally
increased heap sizes of around \(20\%\)~\cite{schism-gc,real-gc-survey}. This
is due to techniques such as fragmented allocation~\cite{bacon-realtime-gc} to
avoid the recompacting stage taken by most non-real-time concurrent collectors
to handle fragmentation.  Fragmented allocation allocates all objects at small,
fixed size chunks, breaking up logical objects larger than the chunk size. The
extra indirection can greatly impact system performance.

\paragraph{Reference counting}
A completely different approach to a tracing garbage collector is reference
counting. Each object has an integer attached to it to count the number of
incoming references, which once it reaches zero, indicates the object can be
freed. It's largely predictable behaviour and simple implementation makes it
common, for example, Python, Objective-C and Swift all use reference counting.

In general, reference counting greatly improves pause times since there is no
background thread for collection, instead reclaiming memory is a incremental
and localised operation. However, three problems emerge: lower throughput,
free-chains and cycles. First, reference counting suffers from poor throughput
due to the need for atomic increments and decrements on pointer modifications.
On average reference counting has \(30\%\) lower throughput compared to tracing
collectors~\cite{blackburn-ulterior,shahriyar-rc}. Recent work has improved
this to be competitive~\cite{shahriyar-rc, shahriyar-rc-immix} but does so by
incorporating techniques from tracing collectors and bringing pauses. Second,
reference counting collectors can suffer from long pauses on free operations
when doing so causes a long chain of decrements and frees to other objects in
the heap. Third, and finally, reference counting suffers from it's inability to
collect cyclic data structures. This is solved by either complicating the
interface to the developer and asking them to break cycles, or by including a
backup tracing collector to collect cycles periodically~\cite{shahriyar-rc}.
Python for example takes this approach.

\section{Design}
\label{sec:design}

\begin{table*}
\centering
\caption{\label{tab:blade_api} \blade\ API}
\begin{tabular}{ll}
  \toprule
  \small{\textbf{Operation}} & \small{\textbf{Description}} \\
  \midrule
  \small{\texttt{void regGCHand\,(handler)}}           & \small{Set function to be called on GC}\\
  \small{\texttt{bool gcHand\,(id, allocated, pause)}} & \small{Upcall to schedule GC event} \\
  \small{\texttt{void startGC\,(id)}}                  & \small{Start the GC} \\
  \bottomrule
\end{tabular}
\end{table*}

\blade\ is an interface to the run-time system (RTS) of the language that
allows programmers to participate in the decision to pause for collection,
letting them customize the collection policy to their system. \blade\ is not a
new approach to garbage collection, but a new approach to dealing with its
performance impact in distributed systems.

Table~\ref{tab:blade_api} summarises the API for \blade, which consists of
three simple functions. The \texttt{regGCHand\,(handler)} simply setups a
function as a target for an upcall from the RTS\@. The \texttt{startGC\,(id)}
starts the collector, passing in an \texttt{id} number previously given to the
application through an upcall. The \texttt{id} argument is a monotonically
increasing argument over upcalls and serves to make the function idempotent.
Finally, the upcall function \texttt{gcHand\,(id, allocated, pause)}, is
invoked by the RTS at the start of every collection and allows the application
developer to decide if the collection should occur immediately or be delayed.
The \texttt{id} argument identifies this collection event, while the \texttt{a}
argument indicates the current heap size. Finally, the \texttt{p} argument
gives an estimate by the collector on the time that this collection will take.
The function can either return a boolean result of \texttt{true} to indicate
that the RTS should immediately perform the collection, or it can return
\texttt{false} to delay the collection until \texttt{startGC} is called.

This API is simple enough that most garbage collected languages can implement
it in a hundred lines of code or so. For example, it took 112 lines of code to
implement \blade\ for the Go programming language.

While there are a few different design choice for the API, we decided on this
one as it is minimal and easily supported by languages, yet expressive enough
for supporting our end-to-end systems.
The parameters passed through to the \texttt{gcHand} function are where we had
the most choice, and indeed the right choice here will likely vary slightly
from language to language.
For exampe, in Java, a third parameter of the amount
of heap remaining would be appropriate, but our target language of Go doesn't
support any notion of bounding the heap size.
The purpose of the arguments to \texttt{gcHand} is to allow the application
developer to make appropriate policy decisions on when to collect.
This will generally be a binary choice of collecting now, or deferring
collection until appropriate failure recovery actions have been taken to
minimize the latency impact.
The right decision for the application will depend on the expected latency
impact of the collection as short collection do not make sense to coordinate
globally.
The estimated pause time in our current Go implementation is derived by simple
linear extrapolation from previous collection pause times at different heap
sizes.

As \blade\ allows delaying collection, the RTS must decide both when to make
the upcall to the application and what to do if memory is exhausted before the
collection is scheduled. For the first situation we add a configurable
low-water-mark parameter to the RTS to allow specifying how much room for delay
should be left when upcalling into the application. For the exhaustion
situation, we simply have the collector run immediately. Any future call to
\texttt{startGC~(id)} with that collection events \texttt{id} number will be
ignored. This retains safety and simply reduces performance in the worst case
to one without \blade. We initially tried adding a second upcall from the RTS
to the application to notify them when this timeout occured, but found that it
was both complex to handle and generally of little benefit. Given that this is
also expected to be rare, moving \texttt{startGC~(id)} to be idempotent
resulted in what we believe to be a stronger design.

\section{\blade\ Systems}
\label{sec:apps}

In this section, we apply \blade\ to two different end-to-end distributed
systems. First we look at the simplest case for \blade, a cluster of stateless
HTTP servers behind a load-balancer, next, we look at the Raft~\cite{raft}
consensus algorithm.

\subsection{HTTP Proxy: No shared state}
\label{sec:app_http}

The most natural application domain for \blade is a fully replicated service
where any server can service a request.
Here we consider a load-balanced HTTP service where a single coordinating
load-balancer proxies client requests to many backend servers.
Typically, all backend servers are identical and the load-balancer uses simple
round-robin to schedule requests.
The load-balancer can also detect when backend servers fail by imposing a
timeout on requests. However, since some HTTP requests might take a while to
service, the load-balancer cannot easily distinguish between a misbehaving
server servicing a fast request, and a properly behaving server servicing a
slow request.  As a result, timeouts are typically set high -- for example, in
the NGINX web server~\cite{nginx}, the default timeout is 60 seconds.

The HTTP load-balanced distributed system has a few unique properties. First,
each request can be routed to any of the replicas. Second, any mutable state is
either stored externally (e.g., in a shared SQL database) or is not relevant
for servicing client requests (e.g., performance metrics). Third, the HTTP
load-balancer acts as a single, centralized coordinator for all
requests\footnote{Some deployments have multiple HTTP load-balancers,
themselves load-balanced with DNS or IP load-balancing, however, commonly each
load-balancer in this case manages a separate cluster anyway to make more
effective load-balancing decisions.}. These three properties make \blade\ easy
to utilize.

The approach is to have a HTTP server explicitly notify the load-balancer when
it needs to perform a collection, and then wait for the load-balancer to
schedule it. Once the collection has been scheduled, the load-balancer will not
send any new requests to the HTTP server, and the HTTP server will finish any
outstanding requests. Once all requests are drained, it can start the
collection, and once finished, notify the load-balancer and begin receiving new
requests.
In most situations, the load-balancer will schedule a HTTP server to collect
immediately. However, it may decide to delay the collection if a critical
number of other HTTP servers are currently down for collection. This allows
the load-balancer to make decisions with throughput impacts in mind.
Figure~\ref{fig:http_code} shows the pseudocode for how a backend server uses
\blade. One subtlety is when deciding to handle a collection, the application
starts a new thread (a cheap operation in Go) as the thread that invoked the
callback is another application thread that just tried to allocate, so may be
holding locks.

\begin{figure}
\begin{lstlisting}[language=go]
func bladeGC(id)
  rpc(controller, askGC)
  waitTrailers()
  startGC(id)
  rpc(controller, doneGC)

func handGC(id, allocd, pause) bool
  if threshold(allocd, pause)
    return true
  else
    // start in new thread
    go bladeGC(id)
    return false
\end{lstlisting}
\caption{\label{fig:http_code} Pseudocode (Go) for HTTP Cluster using \blade.}
\end{figure}

\subsubsection{HTTP: Performance}
\label{sec:http_model}

\begin{figure}
\hspace*{-10mm}
\centering
\begin{tikzpicture}[scale=1.3,align=center]
  \draw[->] (0,0) -- (0,3);
  \draw[->] (0,0) -- (4.7,0);

  \draw (-0.5,1.4) node[rotate=90] {Server Capacity};

  \draw[-] (-0.2,2.8) -- (0.0,2.8);
  \draw (-0.6,2.8) node {100\%};
  \draw (-0.6,0.0) node {0\%};

  \draw[-] (0.5,0) -- (0.5,-0.2);
  \draw (0.5,-0.6) node {can \\ gc?};

  \draw[-] (1.25,0) -- (1.25,-0.2);
  \draw (1.25,-0.6) node {gc \\ ok};

  \draw[-] (2,0) -- (2,-0.2);
  \draw (2,-0.6) node {start \\ gc};

  \draw[-] (3.5,0) -- (3.5,-0.2);
  \draw (3.5,-0.6) node {end \\ gc};

  \draw[-] (4.25,0) -- (4.25,-0.2);
  \draw (4.25,-0.6) node {gc \\ done};

  \draw[dashed,-] (0,2.8) -- (1.25,2.8);
  \draw[dashed,-] (1.25,2.8) -- (2.0,0);
  \draw[dashed,-] (4.25,0) -- (4.25,2.8);
  \draw[dashed,->] (4.25,2.8) -- (4.7,2.8);


  \draw[-] (0.43,-1.0) -- (0.43,-1.2);
  \draw[-] (1.2,-1.0) -- (1.2,-1.2);
  \draw[-] (0.43,-1.2) -- (1.2,-1.2);
  \draw[-] (0.825,-1.2) -- (0.825,-1.3);
  \draw (0.78,-1.6) node {\(t_{schedule}\)};

  \draw[-] (1.3,-1.0) -- (1.3,-1.2);
  \draw[-] (1.95,-1.0) -- (1.95,-1.2);
  \draw[-] (1.3,-1.2) -- (1.95,-1.2);
  \draw[-] (1.615,-1.2) -- (1.615,-1.3);
  \draw (1.8,-1.6) node {\(t_{trailers}\)};

  \draw[-] (2.05,-1.0) -- (2.05,-1.2);
  \draw[-] (3.45,-1.0) -- (3.45,-1.2);
  \draw[-] (2.05,-1.2) -- (3.45,-1.2);
  \draw[-] (2.75,-1.2) -- (2.75,-1.3);
  \draw (2.75,-1.6) node {\(t_{gc}\)};

  \draw[-] (3.55,-1.0) -- (3.55,-1.2);
  \draw[-] (4.25,-1.0) -- (4.25,-1.2);
  \draw[-] (3.55,-1.2) -- (4.25,-1.2);
  \draw[-] (3.9,-1.2) -- (3.9,-1.3);
  \draw (3.9,-1.6) node {\(t_{rpc}\)};

\end{tikzpicture}
\caption{\label{fig:http_cap} Capacity of a HTTP server over time during a
garbage collection cycle with \blade.}
\end{figure}
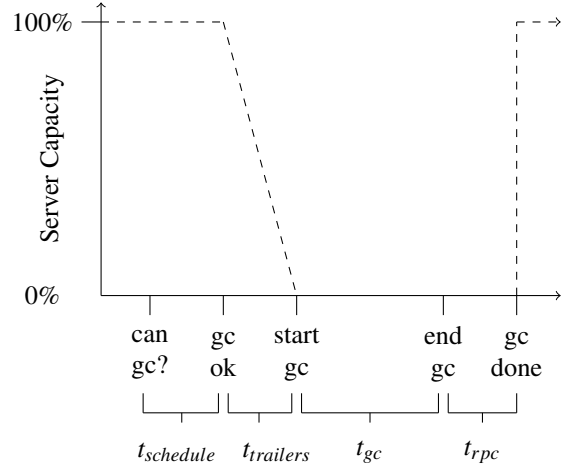

Using \blade\ with the HTTP cluster allows us to trade capacity for better
latency, as such, no request should ever block waiting for the garbage
collector.
We can model this formally to investigate the impact of a GC event on the
system. We break down the stages involved at a single HTTP backend for
performing a garbage collection using \blade; this can be seen in
Figure~\ref{fig:http_cap}. It consists of \(T_{schedule}\), the time to both
request and be scheduled to GC by the load-balancer, \(T_{trailers}\), the time
for the HTTP server to service any outstanding requests, \(T_{gc}\), the time to
perform the garbage collection, and \(T_{rpc}\), the time to send an RPC
notifying the load-balancer the GC is finished. This gives us the following
model:

\begin{center}
\vspace{1em}
\begin{tabular}{ll}
  \toprule
  \multicolumn{2}{c}{\small{\textbf{HTTP Cluster GC Model}}} \\
  \midrule
  \(Latency Impact\)    & \(= 0\) \\
  \(Capacity Loss\)     & \(= 1\ server\) \\
  \(Capacity Downtime\) & \(= T_{trailers} + T_{gc} + T_{notify}\) \\
  \(Event Time\)        & \(= T_{schedule}  + CapacityDowntime\) \\
  \bottomrule
\end{tabular}
\vspace{1em}
\end{center}

In general, we expect \(T_{schedule}\) to be \(1\) network round-trip-time (RTT),
while \(T_{notify}\) should be {\footnotesize\(\frac{1}{2}\)} the network RTT\@.
The value of \(T_{trailers}\) is application specific, but importantly, is a term
expressed in units that the application developer is intimately familiar with.

The latency impact of zero is of course only true when the current throughput
demand on the cluster is low enough to be satisfied by the remaining servers
without queuing. However, even when this isn't the case as the load-balancer
spreads all requests evenly over the remaining servers, no individual request
experiences a disproportionate latency impact. Without \blade, the latency
impact on requests of garbage collection would be the length of the GC pause,
\(T_{gc}\), potentially far longer than \(0\). On the downside, using \blade\
does extend the duration of the capacity downtime by \(T_{trailers} +
T_{notify}\), which has a lower bound of half the RTT.

Importantly this model show how \blade\ allows developers to achieve the three
goals we started with: bounding latency, do so using failure recovery
mechanisms present in the system, and model the performance impact of garbage
collection on the system. For a HTTP cluster, \blade\ bounds latency to 0 by
using the load-balancer and allows us to model this without concern for
workload, heap size or the underlying garbage collection algorithm.

\subsection{Raft: Strongly consistent replication}
\label{sec:app_raft}

In the HTTP load-balancer, because there is no shared mutable state at the
server, any server can service any request and, as a result, we can treat
garbage collection events as temporary failures. The same is true when mutable
state is consistently shared between all servers, for example, as in a
Paxos-like~\cite{paxos} system that uses a consensus algorithm for strongly
consistent replication. In this section we consider how to use \blade\ for the
Raft~\cite{raft} consensus algorithm.

In Raft, during steady-state, all write requests flow through a single server
referred to as the `leader'. Other servers run as `followers'. Writes are
committed within a single round-trip to a majority of the other servers,
leading to sub-millisecond writes in the common case\footnote{In a low-latency
network topology and persistent storage (such as flash drives).}. Garbage
collection pauses can hurt cluster performance in two cases. First, when the
leader pauses, all requests must wait to be serviced until GC is complete. If
GC pause time exceeds the leader timeout (typically 150ms), the remaining
servers will elect a new leader before GC completes. Second, if a majority of
the servers are paused for GC, no progress can be made until a majority are
live again. The second case is worse, because if garbage collection pauses are
very long, there is no built-in way for the system to make progress during this
time. The probability of this occurring is higher than expected as the memory
consumption will be roughly synchronized across servers because of the
replicated state machine each on is executing.

\begin{figure}
\begin{lstlisting}[language=go]
// run in own thread
func bladeClient()
  reqInFlight := 0
  forever
    select
      case id := <- gcRequest:
        reqInFlight = id
        rpc(leader, askGC)

      case id := <- gcAllowed:
        reqInFlight = 0
        startGC(id)
        rpc(leader, doneGC)

      case leader := <- leaderChange:
        if reqInFlight != 0
          rpc(leader, askGC)

func handGC(id, allocd, pause) bool
  if threshold(allocd, pause)
    return true
  else
    // start in new thread
    go func() { gcReq <- id }()
    return false
\end{lstlisting}
\caption{\label{fig:raft_client_code} Pseudocode (Go) for Raft server, when
functioning as a follower and not a leader, using \blade. The \(\leftarrow\)
symbols represent message passing between threads using channels.}
\end{figure}

We use \blade\ with Raft as follows. First, when a follower needs to GC, we
follow a protocol similar to the HTTP load-balanced cluster. The follower
notifies the leader of it's intention to GC and waits to be scheduled. The
leader schedules the collection as long as doing so will leave enough servers
running for a majority to be formed and progress made. We only consider servers
offline due to GC for this, as servers down for other reasons could be down for
an arbitrary amount of time. The leader must also timeout servers considered
down for garbage collection to prevent blocking, marking their GC as completed,
in the rare event that they become unavailable during a collection. One
important difference with Raft compared to the HTTP load-balancer is that the
leader doesn't need to stop sending requests to followers while they are
collecting, and neither do followers need to wait to finish any outstanding
requests. As Raft is designed to make progress with servers unavailable, we can
rely on this and have a follower proceed with GC\@. We present the pseudocode
for the follower situation in Figure~\ref{fig:raft_client_code}, including the
retry logic for sending requests to the new leader if it changes over the
course of a collection request. The code makes use of \emph{channels}, a
message passing mechanism provided in Go.

The second situation, when a server is acting as leader for the cluster, is
more interesting. Since the cluster cannot make progress when the leader is
unavailable, we switch leaders before collecting. Once the leadership has been
transferred, the old leader (now a follower) runs the same algorithm as
presented previously for followers in Figure~\ref{fig:raft_client_code}. A
leadership switch like this can be done in just {\footnotesize\(\frac{1}{2}\)}
the RTT of the network by having the current leader send a broadcast to all
servers in the cluster notifying of the new leader~\cite{raft-thesis}. The
current leader may need to delay switching leadership until it knows that the
next chosen leader is up-to-date, but during this time, the cluster can
continue servicing requests. We present the Pseudocode for the leader situation
in Figure~\ref{fig:raft_leader_code}. In this design the current leader chooses
the last server that collected to be the next leader, or a random server if
this information isn't known. Since the current leader acts as the coordinator
for garbage collection, it also keeps track of how many servers are currently
collecting, and queues requests for future collections from servers that cannot
be scheduled immediately.

Finally, outstanding client requests at the old leader must be handled. One
method is to notify clients of the new leader and have them retry. This is
simple but incurs more latency than required. Instead, the old leader can act
as a proxy for these requests, forwarding them to the new leader in the same
RPC as the election switch message. The new leader can either reply to clients
through the old leader, or directly to them, depending on the client design.

\begin{figure}
\begin{lstlisting}[language=go]
// run in own thread
func bladeLeader() void
  used    := 0
  pending := queue.New()
  lastGC  := cluster.RandomServer()
  forever
    select
      case from := <- gcRequest:
        if used + 1 >= quorum
          pending.End(from)
        else
          used++
          if from == myID
            switchLeader(lastGC)
          else
            rpc(from, allowGC)
    
      case from := <- gcFinished:
        lastGC = from
        if pending.Len() == 0
          used--
        else
          from = pending.Front()
          if from == myID
            switchLeader(lastGC)
          else
            rpc(from, allowGC)

      case leader := <- leaderChange:
        used = 0
        pending.Clear()
}
\end{lstlisting}
\caption{\label{fig:raft_leader_code} Pseudocode (Go) for Raft server, when
functioning as a leader, using \blade.}
\end{figure}

\subsubsection{Raft: Performance}
\label{sec:raft_model}

As before with the HTTP cluster, we can model the performance impact on Raft of
a GC event when using \blade. First, we model the impact when a follower
collect, and secondly, when the leader collects.

In the first case, when a follower collects, the Raft cluster can service this
GC without any impact on the latency of the system. Throughput should also be
unaffected, although we are making the assumption that the cost to bring a
unavailable server up-to-date after a GC does not noticeably affect the
throughput and latency of the cluster. This gives us the model below for the
impact of a follower GC event, where we expect \(T_{schedule}\) to be the
network RTT in the common case:

\begin{center}
\vspace{1em}
\begin{tabular}{ll}
  \toprule
  \multicolumn{2}{c}{\small{\textbf{Raft Follower GC Model}}} \\
  \midrule
  \(Latency Impact\) & \(= 0\) \\
  \(Capacity Loss\)  & \(= 0\) \\
  \(Event Time\)     & \(= T_{schedule}  + T_{gc}\) \\
  \bottomrule
\end{tabular}
\vspace{1em}
\end{center}

In the second case, when a leader collects, then we will take the additional
cost of a fast leader election and proxying queued client requests to the new
leader. This gives us the model below:

\begin{center}
\vspace{1em}
\begin{tabular}{ll}
  \toprule
  \multicolumn{2}{c}{\small{\textbf{Raft Leader GC Model}}} \\
  \midrule
  \(Latency Impact\) & \(= T_{fastelect} + T_{proxy}\) \\
                     & \(= 1 RTT\) \\
  \(Capacity Loss\)  & \(= 0\) \\
  \(Event Time\)     & \(= T_{fastelect} + T_{proxy} + T_{gc}\) \\
                     & \(= \frac{1}{2} RTT + T_{proxy} + T_{gc}\) \\
  \bottomrule
\end{tabular}
\vspace{1em}
\end{center}

One complication with the leader case, captured by the \(T_{proxy}\) value, is
that the leader needs to both forward any queued requests from clients to the
new leader, and also should inform clients that a leadership change has
occurred. The time it takes to do this, and so for how long the leader should
delay beginning its GC, is highly dependent on the system setup. With a small
number of known clients, the leader can broadcast to them that a new leader has
been elected. With a larger, or unknown number of clients, a proxy layer may be
desirable that clients go through.

\section{Evaluation}
\label{sec:eval}

To evaluate \blade, we used it in two distributed systems, first a HTTP cluster
behind a load-balancer, and second, the Raft consensus algorithm. Both systems
are previously described in Section~\ref{sec:apps}.

For evaluating the performance of the GC system, we use the standard Go garbage
collector since all our systems are written in the Go programming language. We
use Go version 1.4.2, the latest at the time of writing. Go currently uses a
parallel mark-sweep collector, with marking done as a stop-the-world phase and
sweeping done concurrently with the application (mutator) threads. Because this
GC design is far from state-of-the-art (although still very common in modern
languages), we also compare against the ideal case of no garbage collection at
all. We do this by simply disabling Go's garbage collector, so memory is never
reclaimed. Go by default also runs the collector every two minutes if not run
recently in order to give memory back to the operating system. For all of the
evaluations below we disable this as we felt it unfairly favoured \blade\ by
being a explicit source of synchronization.

All experiments were run over a 10\,GbE network, using machines with Intel Xeon
E3{-}1220 4 core CPU's with 64\,GB of RAM and running FreeBSD 10.1. The network
RTT was measured to be \(48\mu s\) on average.

\subsection{HTTP Load-Balancer Performance}
\label{sec:eval_http}

In this section we investigate the performance impact of using \blade\ with a
HTTP load-balanced cluster.

We built a simple web application that allows users to search and retrieve
movie information from a backing SQL database. The web app keeps an in-memory
local cache of recent movie insertions and retrievals to improve performance by
avoiding a DB lookup on each requests. The application does not allow updates
to existing records. We run HAProxy~\cite{haproxy} version 1.5 (latest at time
of writing) in front of three servers, using round-robin to load balance
requests across all three.

\begin{table}
\centering
\begin{tabular}{lc}
  \toprule
  \small{\textbf{Component}} & \small{\textbf{SLOC}} \\
  \midrule
  Web Application            &  54 \\
  Load-balancer Coordinator  & 174 \\
\end{tabular}
\caption{\label{tab:http_code} Source code changes needed to utilize
  \blade\ with a web application cluster.}
\end{table}

Adding support for \blade\ to the web application required 228 SLOC to be
added. Of these, 54 were added to the web application itself, while the other
174 were for implementing a controller for the load balancer to coordinate the
GC at each web application server and ensure only one was ever collecting at
any point in time. As HAProxy already supports a TCP interface for enabling and
disabling backends, the coordinator is only required when enforcing capacity
SLAs.

\begin{figure}[t]
  \centering
  \hspace*{-5mm}
  \includegraphics[width=0.5\textwidth]{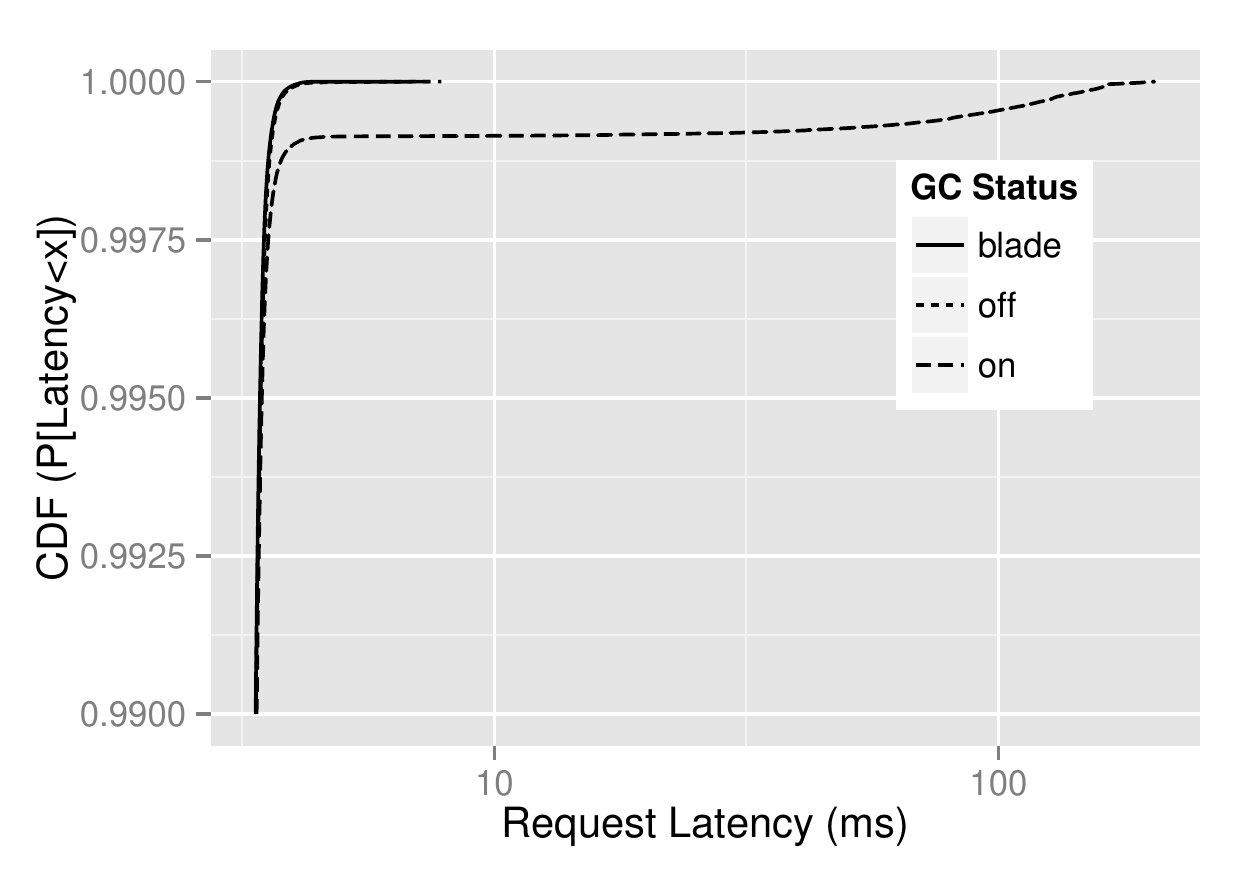}
  \caption{\label{fig:http_latency}
    CDF of request tail-latency to 3-node HTTP cluster behind a load-balancer.
    \blade\ and the \texttt{GC-Off} configuration are so similar that their
    lines overlap.
    Load generator is simulating 400 connections to the cluster,
    sending 6,000 requests-per-second. Each node has a 1GB heap for a 150MB
    live set, and allocates on average at \(12.5\,\text{MB/s}\). Each node
    collects 3 times during the 6 minute experiment.}
\end{figure}

We also evaluated the latency behaviour of the three different configurations
of the cluster using a fifth machine to generate load. A CDF of the request
tail-latency when generating 6,000 requests-per-second can be seen in
Figure~\ref{fig:http_latency}. We ran the experiment for six minutes, during
which each node collects three times. We ran the experiment four times in total
for each configuration and averaged the results. \blade\ achieves a result so
similar to the \texttt{GC-Off} configuration that we have to present them on
the same line in Figure~\ref{fig:http_latency}. Overall performance of each
configuration can be seen in Table~\ref{tab:http_performance}. The
\texttt{GC-On} configuration has tail-latencies far beyond the time the
application is paused by the garbage collector. This appears to be due to the
impact of queues building up, occasional network retransmissions when buffers
overflow, and unfair servicing of pending sockets by Go. This amplification
affect has previously been explored~\cite{tales-of-tail, lightning-cloud}.

\begin{table}
\centering
\begin{tabular}{lrrrr}
  \toprule
  & \small{\textbf{GC-Off}} & \small{\textbf{\blade}} & \small{\textbf{GC-On}} \\
  \midrule
  Mean           & 2.312 &  2.311 &   2.403 \\
  Median         & 2.296 &  2.294 &   2.297 \\
  Std. Dev.      & 0.579 &  0.582 &   3.395 \\
  Max            & 7.847 &  7.443 & 164.206 \\
  Avg. GC-Pause  &     0 & 12.423 &  12.339 \\
\end{tabular}
\caption{\label{tab:http_performance}
  Latency measurements of requests to 3-node HTTP cluster behind a
  load-balancer under different GC configurations. Timings are in milliseconds
(ms). Same experiment as Figure~\ref{fig:http_latency}.}
\end{table}

\begin{figure}[t]
  \centering
  \hspace*{-5mm}
  \includegraphics[width=0.5\textwidth]{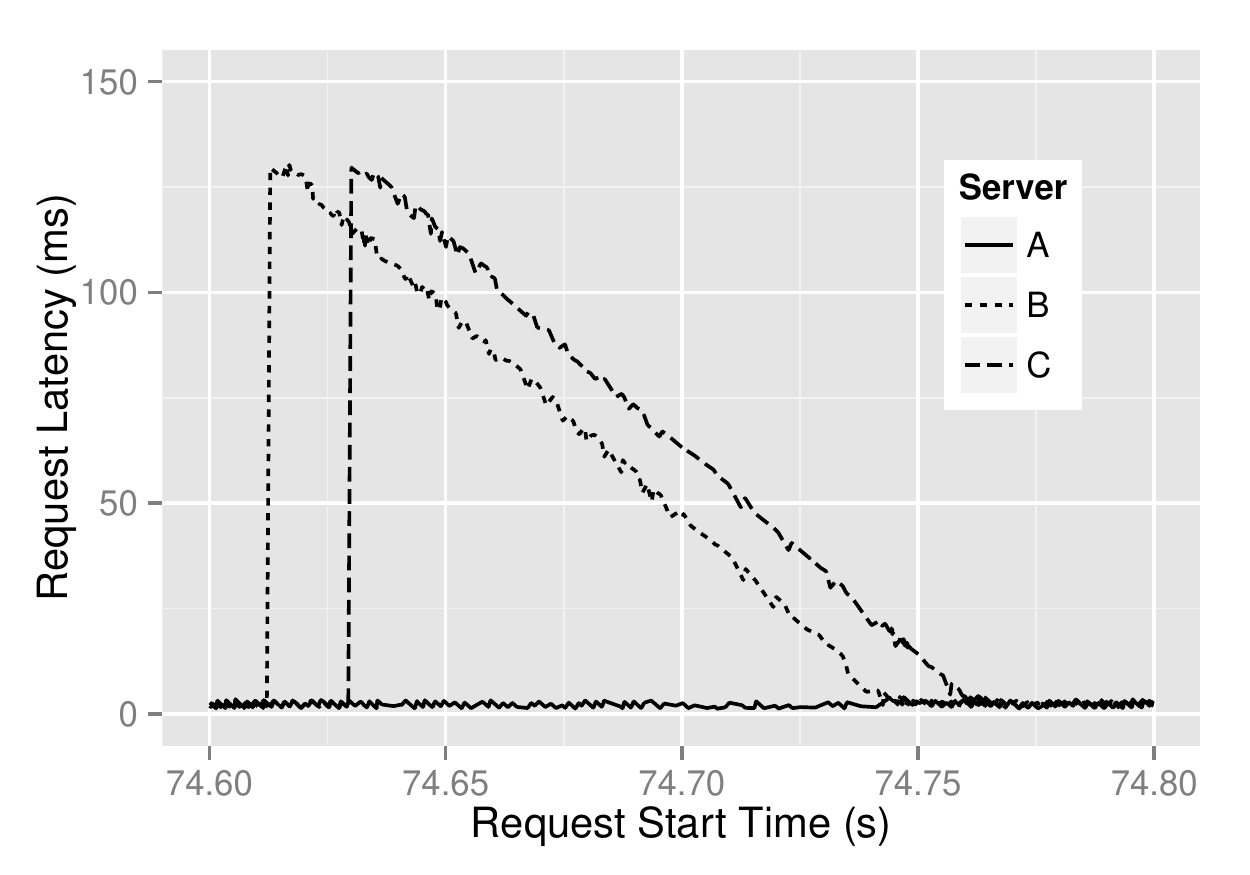}
  \caption{\label{fig:http_gc_overlap}
    Request latency of HTTP cluster broken out by backend server during a
    collection event at two of the workers. Over four runs of the latency
    experiment we observed 36 collections, with 8 of them overlapping, or
    \(22.2\%\).}
\end{figure}

During these runs we also observed occasions when the garbage collection event
at a backend server overlapped with another. An example of such an overlap can
be seen in Figure~\ref{fig:http_gc_overlap}, with the latency of requests to
each server shown as the GC event occurs at servers \(B\) and \(C\). Out of a
total of 36 observed collections across the three servers, 8 of them overlapped
for an average of \(22.2\%\) of collections. While this is likely high due to
the experimental setup, real-world systems often have external sources of
synchronization that increase the chances of these overlaps occurring. For
example, the Go default GC policy of running every two minutes (which we
disabled), or when sudden surges of traffic hits the cluster.

\begin{figure}[t]
  \centering
  \hspace*{-5mm}
  \includegraphics[width=0.5\textwidth]{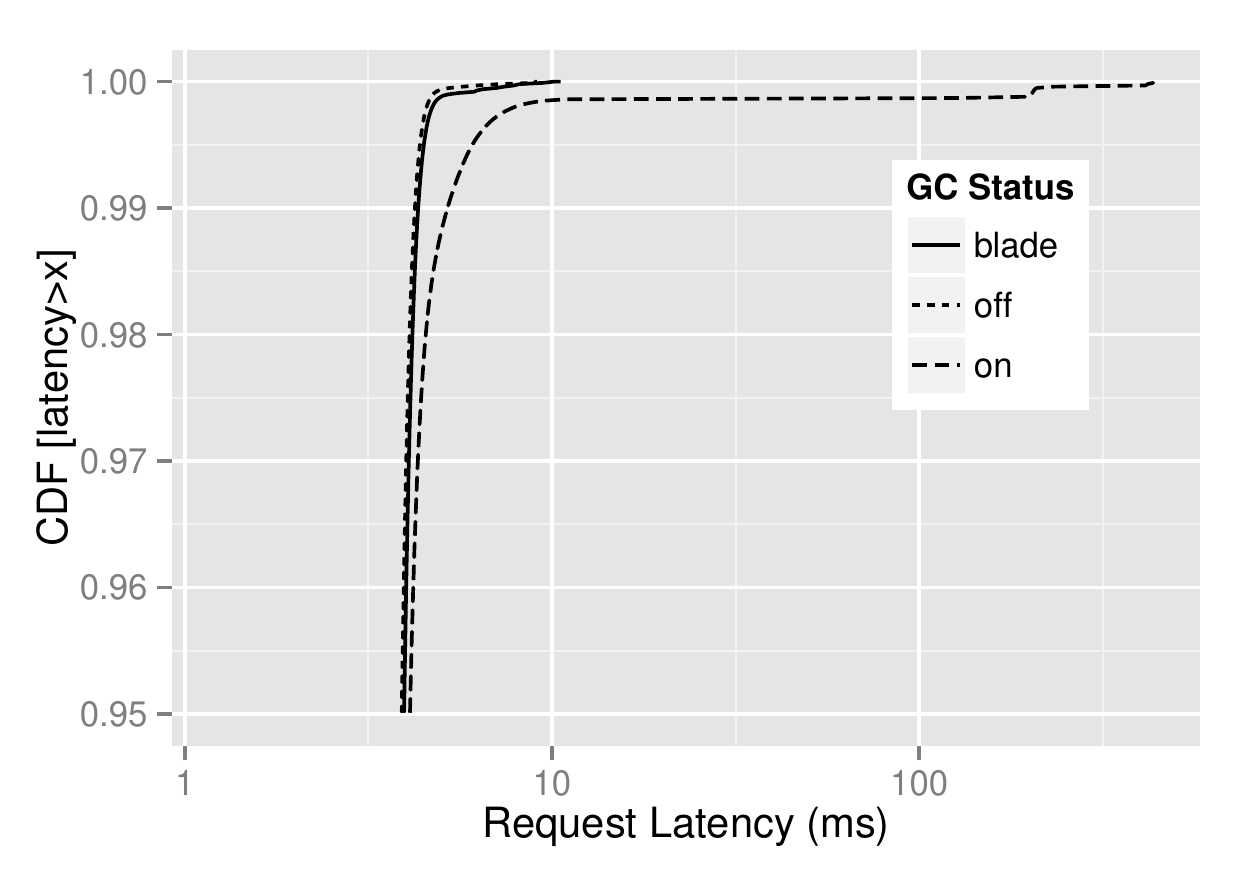}
  \caption{\label{fig:http_latency_40k}
    CDF of request tail-latency to 3-node HTTP cluster behind a load-balancer.
    Load generator is simulating 400 connections to the cluster, sending 40,000
    requests-per-second. Each node has a \(1.5\,\text{GB}\) heap for a
    \(200\,\text{MB}\) live set, and allocates on average at
    \(83.3\,\text{MB/s}\). Each node collects 17 times during the 6 minute
    experiment, with an average pause time of \(14.45\,\text{ms}\).}
\end{figure}

Finally, we ran a second experiment on the same cluster to check the throughput
that each configuration is capable of, the results of which are presented in
Table~\ref{tab:http_throughput}. As expected, \blade\ doesn't cause any drop in
throughput compared to the regular \texttt{GC-On} setup, both achieving around
52,000 requests-per-second. The \texttt{GC-Off} configuration however achieves
a lower throughput due to the overhead of constantly requesting fresh memory
from the OS, consuming a \(34\,\text{GB}\) heap by the end of the experiment.
We used these numbers to run one final latency test, but generating 40,000
requests-per-second this time, close to the peak for all three configurations.
The results can be seen in Figure~\ref{fig:http_latency_40k}.  The slight
penalty that the \blade\ configuration pays at the tail, from reduced capacity,
when under load, can be seen when comparing \texttt{GC-Off} with \blade.
\blade\ is on average \(100\)--\(300\,\mu s\) slower from the 95th percentile on.

\begin{table}
\centering
\begin{tabular}{lccc}
  \toprule
  & \small{\textbf{GC-Off}} & \small{\textbf{\blade}} & \small{\textbf{GC-On}} \\
  \midrule
  Requests/s & 42,643 & 51,624 & 51,983 \\
  Std. Dev.  & 2,213  & 2,573  & 672 \\
\end{tabular}
\caption{\label{tab:http_throughput} Max throughput of each configuration for
  the HTTP cluster. Results are averaged from three runs, each run being 6
  minutes long.}
\end{table}

\subsubsection{Web Application Frameworks}

Using \blade\ in a web application is generic enough in nature that we can
package it as a library. To demonstrate this we wrote a Go package that can be
included by any web application that uses the popular Gorilla Web
Toolkit~\cite{go-gorilla}. It's tied specifically to Gorilla because we need to
be able to detect when all trailing requests have completed (or be able to
cancel them if desired). By including this package, any Gorilla web application
that can work with a client session being handled by different servers, can
benefit from \blade.

\subsection{Raft Performance}
\label{sec:eval_raft}

In this section we investigate the performance impact of using \blade\ with the
Raft consensus algorithm. As Raft is not a standalone system, we use
Etcd~\cite{etcd}, a replicated key-value store with a
ZooKeeper~\cite{zookeeper} inspired API that uses Raft for the consensus
algorithm.

To efficiently use \blade\ in Etcd involved implementing support for
fast-leadership transfers, and also handling GC upcalls using the algorithms
outlined in Figure~\ref{fig:raft_client_code} and
Figure~\ref{fig:raft_leader_code}. This required 563 lines of code to be
changed (largely additions) in Etcd, with the breakdown shown in
Table~\ref{tab:etcd_code}. As fast-leadership transfers are useful for purposes
beyond \blade, it is fair to count the effort needed to support \blade\ in Etcd
as \(349\) SLOC\@.

\begin{table}
\centering
\begin{tabular}{lc}
  \toprule
  \small{\textbf{Component}} & \small{\textbf{SLOC}} \\
  \midrule
  Fast-leader switch & 214 \\
  Blade GC support   & 349 \\
\end{tabular}
\caption{\label{tab:etcd_code} Source code changes needed to utilize
  \blade\ in Etcd.}
\end{table}

For evaluating the performance of \blade\ with Raft, we set up a three node
Etcd cluster under three different configurations. First, when running with the
standard Go garbage collector, secondly, when running with the garbage
collector disabled, and finally, when using \blade. We ran a single experiment
were we loaded 600,000 keys into Etcd and then sent 100 requests per second at
regular intervals for 10 minutes to the cluster using a mixture of reads and
writes in a \(3:1\) ratio. We track the latency of each request after the
initial load of keys. We ran the experiment three times for each configuration
and took the average of the three. In all configurations the standard deviation
between the three runs was less than 5\%. We use a low request rate as at this
time, Etcd is early in its development and doesn't support a high request rate,
(peaking at around 400 requests/s on our setup) becomes very unstable anywhere
close to its peak

With the GC enabled, this experiment peaks at consuming \(473\,\text{MB}\) of
memory. While very small by modern server standards, it is sufficient to
evaluate our results since \blade\ thankfully is not affected by heap size in
terms of latency impact on requests.

\begin{table}
\centering
\begin{tabular}{lrrrr}
  \toprule
  & Mean & Median & Std. Dev. & Max \\
  \midrule
  \small{\textbf{GC-Off}} & 0.532 & 0.530 & 0.031 &  1.127 \\
  \small{\textbf{\blade}} & 0.505 & 0.499 & 0.030 &  1.015 \\
  \small{\textbf{GC-On}}  & 0.589 & 0.517 & 2.112 & 95.969 \\
\end{tabular}
\caption{\label{tab:etcd_performance}
  Latency measurements of SET requests to Etcd cluster under different GC
  configurations. Timings are in milliseconds (ms).
}
\end{table}

The results for \texttt{set} request latencies for all three configurations are
shown in Table~\ref{tab:etcd_performance}. Excluding tail-latency, all three
achieve similar performance levels, although \blade\ outperforms each
configuration across the board. \blade\ achieves a mean of \(505\,\mu s\) and a
worst-case of \(1.01\,\text{ms}\), \texttt{GC-Off} a mean of \(532\,\mu s\) and
a worst-case of \(1.13\,\text{ms}\), and \texttt{GC-On} a mean of \(589\,\mu
s\) and a worst-case of \(95.96\,\text{ms}\). The reason \blade\ even
outperforms the \texttt{GC-Off} configuration is due to the penalty
\texttt{GC-Off} pays from the extra system calls and lost locality from
requesting new memory rather than ever recycling it. Results for \texttt{get}
requests show the same relation among the three configurations.

\begin{figure}[t]
  \centering
  \hspace*{-5mm}
  \includegraphics[width=0.51\textwidth]{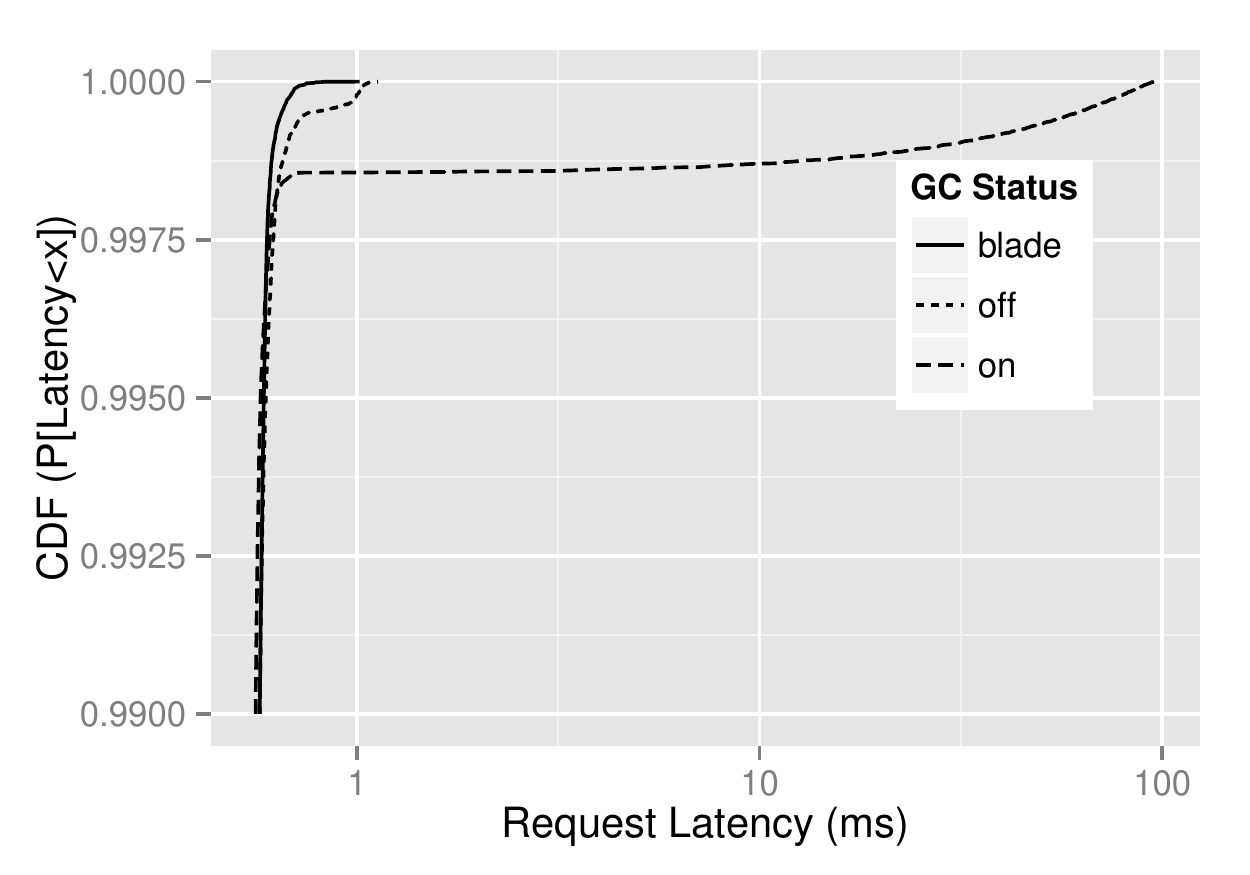}
  \caption{\label{fig:etcd_latency_all}
    CDF of Etcd replicated key-value store request latency for all GC
    configurations}
\end{figure}

When looking at the tail-latency of each configuration, a different story
emerges. A CDF of slowest \(1\%\) of both \texttt{get} and \texttt{set}
requests can be seen in Figure~\ref{fig:etcd_latency_all}. As expected,
performance of the standard GC configuration has a very long tail from GC
pauses. The results for the \texttt{GC-Off} configuration and the \blade\
configuration however are nearly identical. This is expected from the
performance model we established in Section~\ref{sec:raft_model}, that showed
latency has an expected increase of 1 network RTT during a GC, a value of
\(48\,\mu s\) in our experimental setup.  Indeed, as can be seen in the
detailed breakout of the performance in Table~\ref{tab:etcd_latency}, \blade\
slightly outperforms the \texttt{GC-Off} configuration. This is due to the
\texttt{GC-Off} configuration paying a penalty from the higher memory use as
mentioned previously, and the \(48\,\mu s\) being within the tail-latency
caused by other sources of jitter such as the OS scheduler.
 
\begin{table}
  \centering
  \begin{tabular}{lrrr}
    \toprule
    & \small{\textbf{\blade}} & \small{\textbf{GC-Off}} & \small{\textbf{GC-On}} \\
    \midrule
    95th      &  0.52 &   0.54 &  0.54 \\
    99th      &  0.57 &   0.57 &  0.56 \\
    99.9th    &  0.62 &   0.67 & 28.81 \\
    99.99th   &  0.70 &   1.03 & 86.98 \\
    99.999th  &  0.80 &   1.08 & 94.22 \\
    99.9999th &  0.97 &   1.12 & 95.96 \\
  \end{tabular}
  \caption{\label{tab:etcd_latency}
    SLA measurements for different Etcd configurations. Timings are in
    milliseconds (ms).}
\end{table}

\section{Discussion \& Limitations}
\label{sec:discussion}

\paragraph{Understanding Performance}

With \blade\ we set out to achieve three goals: bound tail-latency in
distributed systems, do so using system specific failure recovery mechanisms,
and to allow the performance impact of garbage collection to be modelled
without knowledge of production workloads. For the final point, the performance
models for each system in Section~\ref{sec:design} demonstrate how \blade\ can
achieve this. With Raft for example, we know that the latency impact of using a
garbage collected language will be a single extra network RTT\@. As the time to
complete any request is at least one network RTT, using a garbage collector
with \blade\ limits the tail-latency from GC to twice the mean in the worst
case. Importantly, the impact of GC is now in units comparable to the rest of
the system.  Furthermore, as network speeds improve over time, so will \blade.
While our test setup had a network RTT of \(48\,\mu s\), 10\,Gbps NICs are
currently available that achieve less than \(1.5,\mu s\) latency at the
end-host~\cite{connectx-nic}. Garbage collectors are chasing a continually
moving target, but \blade\ scales with the performance of the distributed
system.

\paragraph{Limitations}

Another important outcome from using \blade\ in a distributed system, is the
changed requirements for the garbage collector. As \blade\ deals with the
latency impact, in most situations a concurrent collector will no longer be the
best fit. Instead, a simpler and high-throughput stop-the-world collector is
best suited~\cite{stw-gc-scale}. These collectors are already readily available
in most languages, unlike high-performance concurrent collectors.

There are, however, a number of limitations with \blade.
\begin{itemize}
  \item First and foremost, \blade\ is not a universal solution. We
    specifically target distributed systems as this is an important area where
    low-latency matters and we've had bad experiences with garbage collected
    languages. Even then, \blade\ wil not work for every distributed system as
    it relies on their being a failure recovery mechanism that can be
    exploited. Common, but not universal.

  \item Secondly, \blade\ requires developers to write code and doesn't apply
    transparently. This, however, is by design and we believe our results show
    that the amount of work needed is low. Even with garbage collection,
    developers do not ignore memory management and still apply techniques such
    as local allocation caches to improve performance, \blade\ is one more
    technique that can be used.

  \item Third, \blade\ takes whole servers offline during a garbage collection,
    which may be too large a capacity loss for some systems. If, for example,
    \blade\ was used for a HTTP cluster with only two servers, then using 50\%
    of the cluster capacity is likely to be unworkable.
\end{itemize}

\section{Related Work}
\label{sec:related}

\paragraph{Trash Day}

Mass et al.\ have recently done work on coordinating garbage collection in a
distributed system~\cite{mass-hotos}. They look at two different systems,
Spark~\cite{spark} and Cassandra~\cite{cassandra}, noting that for Spark,
having all nodes collect at the same time improves performance, while for
Cassandra, staggering collection and routing requests around nodes can reduce
tail latency. They design a run-time system to provide a general approach to
this problem, allowing different coordination strategies across multiple nodes
to be implemented.

\paragraph{Process Restarts}

We have heard of a few different companies in industry that disable garbage
collection, either completely or for the old generation, and then kill and
restart the process as needed. They will often attempt to drain requests before
restarting the process. This is similar to \blade\ but less principled, support
is not provided directly in the programming language and as it requires that
programs can support arbitrary restarts, it only works for a subset of the
programs that \blade\ supports. Restarting should also be a slower operation as
it needs to reload state from permanent storage. As far as we are aware, none
of these companies apply this technique to stateful systems such as Raft.

\paragraph{HTTP Load-balancing}

Portillo-Dominguez et al.\ have recently done work on HTTP load-balancers in
Java to avoid the impact of garbage collection on latencies~\cite{http-gc1,
http-gc2}. Their approach is very similar to \blade, modifying a round-robin
routing algorithm to avoid the collecting server. They do not modify the
language or RTS however as we propose, instead they model the GC and try to
predict when it will collect. Mispredictions mean lower performance than
\blade, and also no ability to deal with overlapping collections. They also
deal with a very different level of performance than we are concerned with,
starting with worst case request latencies in the hundreds of seconds and
reducing that to the tens of seconds. We are instead concerned with
microseconds.

\paragraph{JVM \&~.NET}

The Java Virtual Machine (JVM) supports two programmable interfaces to the
garbage collector. One is the \texttt{System.gc()} function that suggest to
the RTS to start the GC\@. The other is the \emph{Garbage Collection
Notifications API} (JGCN) optional extension~\cite{java-gc-notifications}. JGCN
supports callbacks, like \blade, to the application, but it only supports
notifying the application after a collection has complete. JGCN is intended for
performance monitoring and debugging.

Microsoft's~.NET platform supports an API very similar to \blade, the
\emph{Garbage Collection Notifications} API (MGCN)~\cite{clr-gc-notifications}.
MGCN, like \blade, supports application callbacks before garbage collection
occurs. MGCN doesn't, however, allow the application to delay collection, only
to start it earlier than the RTS planned. MGCN is suggested for use by
Microsoft in a similar manner to \blade, but as of this time we are unaware of
any reports on it's usage or evaluation of it. The lack of control with MGCN to
coordinate nodes, and avoid GC overlaps at servers, appears to be a concern
with some potential users. The popular Stack Overflow website, for example,
chose not to use MGCN partially for this reason~\cite{stackoverflow-gc}.

\paragraph{Concurrent Tracing Garbage Collectors}

A vast amount of work has been done in improving pause times of garbage
collectors. A sample of this was covered in Section~\ref{sec:background}. Azul
Systems Zing GC~\cite{pauseless-gc, c4-gc, collie-gc, azul-white-paper1} is one
of the best available today, with pause times in the low milliseconds or
microseconds. This is still one to two orders of magnitude above what \blade\
can achieve, and will get worse as faster networks with under \(5\,\mu s\) RTT
become available. The work by Pizlo et al.\, such as Schism, on real-time
collectors~\cite{schism-gc, real-gc-survey} achieves the lowest pause times we
are aware of, capable of bounds in the tens of microseconds but suffers from
30\% lower throughput and 20\% higher memory consumption compared to
stop-the-world collectors. Other concurrent and real-time collectors that we
are aware of~\cite{mapping-gc, compressor-gc, bacon-realtime-gc, g1-gc,
sapphire-gc, cheng-gc, staccato-gc, syncopation-gc} all perform worse in
latency and/or throughput than both Azul and Schism, with pause times in the
tens of milliseconds. Finally, while many of these systems have STW pauses for
some situations, work such as that of Tomoharu et al.~\cite{references-gc}
seeks to address these final cases.

\paragraph{Reference Counting}

The best reference counting collectors have very low and uniform latency impact
on an application as demonstrated by the Ulterior Reference
Counting~\cite{blackburn-ulterior} collector. However, they have historically
suffered from lower throughput compared to tracing collectors. The work of
Shahriyar et al.~\cite{shahriyar-rc, shahriyar-rc-immix} has made reference
counting collectors competitive, but does so by incorporating background tasks
and pauses. Unfortunately Shahriyar doesn't report the latency impact of these
changes.

\paragraph{Tail tolerance}

Vulimiri et al. proposed~\cite{more-is-less} an approach to handling
tail-latency for Internet services such as DNS, where requests were duplicated
and sent to multiple servers. Dean and Barroso proposed and investigated a
similar idea~\cite{tail-at-scale}, but specifically for addressing
tail-latency~\cite{tail-at-scale} in data centers. Servers then either race to
fulfil the request, or coordinate with each other to claim ownership of the
request when they start processing it. Jalaparti applied this idea, as well as
allowing incomplete requests, to build a framework for constructing data center
services~\cite{speeding-req-resp}. \blade\ takes a similar approach to tail
tolerant systems, not attempting to reduce the impact of garbage collection on
an individual server, but avoiding it's impact on the end-to-end system.
\blade\ however solves a specific, but common problem, garbage collection,
rather than treating servers as a black box. This allows \blade\ to be used in
situations such as Raft where tail tolerant systems do not apply as requests
cannot be duplicated and sent to multiple servers.

\section{Conclusion}
\label{sec:conclusion}

\blade\ is a new approach to garbage collection for a particular, but large and
important class of programs: distributed systems. \blade\ uses the ability of
distributed systems to deal with failure, to also handle garbage collection,
treating garbage collection as a partially predictable and controllable
failure.
We applied \blade\ to two important and common systems, a cluster of web
servers and the Raft consensus algorithm. For the first case, we eliminated the
latency impact of garbage collection, and for the second, we reduced it to the
order of a single network round-trip, or \(48\,\mu s\) in our experiment.
As \blade\ handles the impact of garbage collectors in distributed systems
rather than attempt to improve them directly, it allows for a different set of
choices when designing the collector. Simple, high-throughput designs are
preferable with \blade\ than the complexity of collectors that try to minimize
pause times.

\section*{Acknowledgments}

We thank Keith Winstein for his help in reviewing drafts of this paper.

{%
  \frenchspacing
  \bibliographystyle{acm}
  \bibliography{paper}
}

\end{document}